\documentclass[xypic,amscd,syntonly,amssymb,verbatim,12pt]{amsart}

\usepackage{graphicx}
\usepackage[all]{xy}
\usepackage{amssymb}
\usepackage{amsmath}
\usepackage[mathscr]{euscript}

\usepackage{epsfig}
\theoremstyle{plain}

\newtheorem{Prop}{Proposition}
\newtheorem{Cor}{Corollary}
\newtheorem{Lem}{Lemma}
\newtheorem{Def}{Definition}

 \theoremstyle{definition}
\theoremstyle{remark}

\numberwithin{equation}{section}

\begin{document}
  \title[  Gauge fields and the R-H correspondence  ]{Flat gauge fields and the Riemann-Hilbert correspondence}

 \author{ Andr\'{e}s   Vi\~{n}a}
\address{Departamento de F\'{i}sica. Universidad de Oviedo.    Garc\'{\i}a Lorca 18.
     33007 Oviedo. Spain. }
\email{vina@uniovi.es}

 \maketitle
\begin{abstract}

The geometric phase that appears in the effects of Aharonov-Bohm type  is interpreted in the frame of Deligne's version of the Riemann-Hilbert correspondence. We extend also the concept of flat gauge field to $B$-branes on a complex manifold $X$,   so that such  a field on a $B$-brane turns it into an object of the category of constructible sheaves on $X$. 

\end{abstract}

\section {Introduction} \label{S:intro}

We recall some well-known notions about the gauge fields   in order to better explain the  purpose  of the article.  A matter field on a manifold $X$ is a section of a $C^{\infty}$ vector bundle $F\to X$ over $X$. To ``derive'' the sections of $F$ it is necessary 
to introduce a gauge field, which allows us to define a covariant derivative of   sections of $F$ and the corresponding parallel transport.

From the point of view of the {\it classical} Physics, the gauge field has not physical sense. The relevant magnitude is the strength of the 
field.
The   fields whose strength vanishes are called flat   fields. 
When $X$ is an oriented Riemannian compact manifold and $F$ is a Hermitian vector bundle, it is possible to define a norm for the strength of a gauge field over $F$. The functional that associates to each   field the norm of the  corresponding strength 
is the Yang-Mills functional. If there are flat gauge fields over $F$, 
they  minimize this functional; these fields are the vacuum states of the corresponding Yang-Mills theory (see 
Section \ref{S:Flat}).

 A flat gauge field determines through the parallel transport a representation of the homotopy group $\pi_1(X)$. In this way, one has a relation between the set of vacuum states of the Yang-Mills theory and the representations of $\pi_1(X)$.

On the other hand,  the context of the $C^{\infty}$ vector bundles endowed with smooth gauge fields is not large enough to develop some physical theories. 
{\it Singular} gauge fields are  present in the study of some physical systems;
 for instance, by dealing with the topological phases which appear when a {\it quantum} system performs  a cyclic evolution. These phases are manifestations of the singularities of the flat  fields. The best known case is the Aharonov-Bohm phase, that we briefly recall.


  An indefinitely large solenoid creates a magnetic field $\vec B$ confined inside the coil. 
 The Aharonov-Bohm effect occurs when one considers the motion of an electron in the region   outside the solenoid, where the magnetic field vanishes.
 The potential vector for $\vec B$ is 
necessarily   singular. This singularity is related with the phase shift  in the wave function of the electron, after traveling around a closed path. In fact, 
one has a  phase factor belonging to ${\rm U}(1)$ for each loop around the   singularity. 
A similar situation appears in the description of the Aharonov-Casher effect.

 Another case in which the need to take into account more general structures than vector bundles is in dealing with $B$-branes. 
 One can certainly consider    over a Calabi-Yau manifold $B$-branes defined 
by holomorphic vector bundles. But the homological  mirror symmetry conjectured by Kontsevich imposes to consider more general branes; namely, {\it complexes of coherent sheaves} \cite[Section 5.4]{Aspin} \cite[Section 5.3]{Aspin-et}.  A purpose of this note is to extend the concept of flat gauge field to $B$-branes, admitting the possibility that it has singularities. 
 

In mathematical terms, the space of sections of the vector bundle $F$ is a locally free sheaf ${\mathcal F}$, and a  flat gauge field   on $F$ is an integrable connection $\nabla$ on ${\mathcal F}$. This connection   defines  on ${\mathcal F}$ a ${D}$-module structure \cite[page 18]{Hotta}.
 We remark that
the   equation $\nabla s=0$ for the parallel transport of sections of $F$ is a overdetermined differential equation; that is, the space of covariantly constant sections of $F$ is finite dimensional.

The first step in the above-mentioned extension is to define a flat gauge field 
	over a  sheaf 
 which is not locally free. According to the above argument, it seems reasonable
 to define a flat gauge field on such a sheaf
 as a $D$-module structure on it. The  
 over-determination   of the parallel transport equation can be translated to the  context $D$-modules requiring to the $D$-modules to be holonomic \cite[page 133]{{M-S(1)}}. 
The holonomic $D$-modules are ``generically'' integrable connections; that is, if ${\mathcal M}$ is a such $D$-module over $X$, then there exists an open dense subset $U$ of $X$, such that ${\mathcal M}|_U$ is an integrable connection. 

 The sheaves that are $B$-branes on $X$ are ${\mathcal O}_X$-coherent. On the other hand,  a $D$-module   ${\mathcal O}_X$-coherent   is necessarily an integrable connection \cite[page 184]{Gelfand-Manin2} \cite[page 61]{Hotta}. Thus, a $D$-module structure on such a brane is only possible  if it is a    locally free sheaf.

 To avoid this difficulty and taking into account   the existence of singularities in some physical gauge fields, we will admit the presence of mild singularities in the gauge fields on those branes. 
A such  singular integrable connection defines a $D$-module structure on an extension of the sheaf, which in turn determines, through the Riemann-Hilbert correspondence, representations of the homotopy groups
of the strata of a 
stratification of $X$. These representations can be interpreted as ``generalized Aharonov-Bohm   phase factors''. 

The above idea can be extended  in a natural way to   complexes of 
coherent sheaves, i.e. to general $B$-branes. A flat gauge field on the brane defines a complex of $D$-modules that, by   the Riemann-Hilbert correspondence,
determines local systems
which can be also regarded as   phase factors. 

This article is organized as follows:  Section \ref{Sub:Camposgauge} deals with  basic notions related to the fields and their mathematical meanings.
In Section \ref{S:Conn and D-}, we recall some mathematical definitions and results that will be used below. More concretely, we state  properties of integrable connections and explain briefly some points of the Riemann-Hilbert correspondence.

The flat non singular gauge fields on the manifold $X$ determine representations of the group $\pi_1(X)$, and conversely.
Assumed that $X$ is endowed with a Riemannian metric  and $F$ is an ${\rm U}(m)$ flat vector bundle, then 
  the flat gauge fields 
	are the vacuums of the corresponding field theory, as we have said. Using the correspondence between flat gauge fields and representations,
 we  determine, 	in Section \ref{S:Flat},  the space of vacuum states when 
	$\pi_1(X)$ is an infinite cyclic group and when $\pi_1(X)\simeq {\mathbb Z}/2{\mathbb Z}$ (Propositions \ref{pi1=Z} and \ref{pi1=Z2}).

In Section \ref{S:TopologFases},  we  consider the  geometric  phase shifts which appear in the  wave function of a particle that undergoes a cyclic evolution in a singular flat  field. We study the Aharonov-Bohm and the Aharonov-Casher effects, and the case of a particle that carries a like spin variable. 
We relate those phase shifts   to the representations of   homotopy groups, which are associated to  
meromorphic flat connections by Deligne's version of the Riemann-Hilbert correspondence (Proposition \ref{Prop:Wong}).
 Roughly speaking, 
   we  interpret  the  ``Aharonov-Bohm phase'', that occurs in general gauge field theories, in the context of Deligne's theory.

 In Section \ref{S:FlatConnections}, we define flat gauge fields on a coherent ${\mathcal O}_X$-module  and on objects of the bounded derived  category $D^b(X)$ (of coherent sheaves on $X$), developing the idea explained above. Finally, we consider the classical Aharonov-Bohm  effect in the context of perverse sheaves.
 
 \smallskip
 
 {\bf Acknowledgements.} I am grateful to an anonymous referee for his constructive comments, and for pointing out an error in an earlier version of this article.

\section{Gauge fields, Connections and $D$-modules}\label{S:gauge_fields}

\subsection{Gauge Fields}\label{Sub:Camposgauge} 
We revise  briefly some well-known  physical notions relative to the gauge fields, and recall its mathematical translations. 

The value $\psi(x)$ of the  vector field  $\psi$ at the point $x$ of the space-time $X$, is an element of a certain complex vector space $V_x$. 
  If  $f$ is a complex function defined on $X$, the field     $f\psi$ is defined     by  $(f\psi)(x)=f(x)\psi(x)$. 
   
   A observer ${\sf O}$ ``placed'' at a point $x_0$, 
	by means of his measuring devices, 
 can assign 
  coordinates $(\psi^a(x))$ to $\psi(x)$.   That is, the observer fix   bases in the vector spaces $V_x$.
	

If the observer is assigning bases to all the $V_x$ with $x$ varying along a closed curve, when he returns to the starting point, he must arrive with the same reference system he started with. In other  case, the observer's measurements would not be  consistent.
This consistency may not be achieved, if the observer's domain $U$ is not a ``small enough'' 
neighborhood of $x_0.$	It is admitted that an observer placed in the point $x_0$ can fix reference systems for only   the vector spaces $V_x$, with $x$ belonging to a neighborhood  $U$   of $x_0$.

Another observer ${\sf O'}$, whose domain of observation is ${U'}$, will assign to the vector $\psi(x)\in V_x$ with $x\in {U'}$ the coordinates $\left(\psi'^{a}(x)\right)$. On ${U}\cap{U'},$ we will have defined a matrix function $g$ that relates the coordinates assigned to the same vector by both observers; that is, we have the matrix relation
\begin{equation}\label{phi=gphi}
\left(\psi'^{a}(x)\right)=\left(\psi^{a}(x)\right)\cdot g(x).
\end{equation}
Admitting that all vector spaces $V_x$ have dimension $m$ and that   this field  theory has as symmetry group a subgroup $G$ of ${\rm GL} (m, {\mathbb C})$, then the function $g$ takes values in $G$.

We suppose that there exists a set of local observers whose domains form a covering of $X$. Then the set of vector
 spaces $\{V_x\}$ form a {\it vector bundle} ${V}\to X$ over $X$, where the local trivializations are defined by the observers. From the mathematical point of view, the field $\psi$ is a {\it section} of $V$.

 Since $\psi(x)$ and $\psi(y)$ belong to different vector spaces, the expression $\psi(y)-\psi(x)$ does not make sense, so it is not possible to define the derivative of the $\psi$ field as a simple incremental quotient. It is a matter of defining
 $D_v\psi$, the variation of $\psi$ in the direction of a vector $v$ tangent to $X$ at the point $x$; that is, the covariant derivative of $\psi$. 
It seems reasonable to demand that $D$ satisfies the following properties  
\begin{equation}\label{LeibnizRule}
D_v(f\psi)=v(f)\psi+fD_v\psi, \;\;\;D_v(\psi+\phi)=D_v(\psi)+D_v(\phi).
\end{equation}
In mathematical terms, an operator  $D$ satisfying  (\ref{LeibnizRule}) 
is a {\it connection} on the vector bundle $V$.

If an observer fix a local frame $s=(s_1,\dots, s_m)$  on the points of ${U}$, then to define $D$ is necessary to give 
$D_vs_a$ para $a=1,\dots,m$. That is, to express $D_vs_a$  as a linear combination of the $s_b$'s,
\begin{equation}\label{Dvsa}
D_v(s_a)=\sum_bA_{ab}(v)s_b.
\end{equation}
 It   seems reasonable to impose that $D_v$ be linear with respect to $v$. Thus, $A=(A_{ab})$ must be a matix of $1$-forms on $U$.
 
Another observer fix the local frame $s'=(s'_1,\dots,s'_m)$ on $U'$. Then (\ref{phi=gphi}) implies that, on $U\cap U'$,  the following matrix equality holds
\begin{equation}\label{gcdots'}
g\cdot s'=s,  
\end{equation}
where $s$ and $s'$ are written as $m\times 1$-matrices. The frame $s'$ will determine a matrix $A'$ of $1$-forms on $U'$, such that  
 $Ds'_a=\sum_b A'_{ab}s'_b$. From (\ref{gcdots'}) together with (\ref{LeibnizRule}), it follows 
$$A=dg\cdot g^{-1}+g\cdot A'\cdot g^{-1},$$
on the points of $U\cap U'$. That is, the set of $1$-forms $\{A, A',\dots\}$ is a {\it gauge field} on $X$. It determines the covariant derivative $D$, and conversely.

The  operator $D$ can be extended, consistently with the exterior differentiation of forms, to combinations
$\sum \alpha^c s_c\equiv \alpha\cdot s$, where the coefficients $\alpha^c$ are $1$-forms
$$D(\alpha\cdot s) =\left(d\alpha-\alpha\wedge A\right)\cdot s.$$
 In particular, 
  $$D^2s=D(A\cdot s) =: C\cdot s,$$
  where $C$ is the matrix of $2$-formas on $U$  
  \begin{equation}\label{Curvature}
  C=dA-A\wedge A.
  \end{equation}
$C$ is the expression of the {\it strength} of the gauge field on $U$.  On $U\cap U'$ the following relation holds
$$C=g\cdot C'\cdot g^{-1}.$$
 The gauge field is called {\it flat} if its  strength vanishes.

In mathematical terms the set $C, C', \dots$ are {\it curvature} forms of the connection. And  when the curvature vanishes,   the connection is said to be {\it integrable}.


\subsection{Connections and $D$-Modules}\label{S:Conn and D-}

Here, we recall some well known mathematical concepts that formalize the physical notions introduced
 in Section \ref{Sub:Camposgauge}. We also recollect some results which will be used later.

\subsubsection{Integrable connections.}\label{Sub:Flat}
 Let $X$ be a connected complex analytic $n$-manifold, and $F$ a holomorphic vector bundle on $X$. By ${\mathcal F}$ we denote  the ${\mathcal O}_X$-module   of sections of $F$.  
 Denoting by $\Omega^i_X$ the sheaf of $i$-differential forms on $X$, 
  a  connection on ${\mathcal F}$ is a ${\mathbb C}_X$-linear morphism of abelian sheaves
$$\nabla:{\mathcal F}\to  \Omega^1_{X} \otimes_{{\mathcal O}_X} {\mathcal F},$$
satisfying the Leibniz's rule
\begin{equation}\label{Leibniz}
\nabla(f\sigma)=df\otimes \sigma+f\nabla \sigma,
\end{equation}
where $f$ and $\sigma$ are sections of ${\mathcal O}_X$ and ${\mathcal F}$ (resp.) defined on an open subset of $X.$

The connection $\nabla$ can be extended to a morphism
$$\nabla^k: \Omega^{k}_{X}  \otimes_{{\mathcal O}_X} {\mathcal F} \to \Omega^{k+1}_{X}  \otimes_{{\mathcal O}_X} {\mathcal F} ,$$
in the usual way
\begin{equation}\label{nabla_k}
\nabla^k(\alpha\otimes\sigma)=   d\alpha \otimes \sigma +(-1)^k\alpha\wedge\nabla\sigma.   
\end{equation}

The curvature of $\nabla$ is the morphism 
$$K_{\nabla}:=\nabla^1\circ\nabla:{\mathcal F}\to \Omega^2_{X} \otimes_{{\mathcal O}_X}{\mathcal F}.$$
It is well known that $K_{\nabla}$ is ${\mathcal O}_X$-linear \cite[page 3]{Koba}, and that
\begin{equation}\label{integrable}
\nabla^{k+1}\circ\nabla^{k}(\alpha\otimes\sigma)=\alpha \wedge K_{\nabla}(\sigma).
 \end{equation}
The connection $\nabla$ is called integrable (or flat) if $K_{\nabla}=0$.

\smallskip

If $F$ is vector bundle  
of rank $m$ and   $s=(s_1,\dots,s_m)$ be a local frame of $F$ defined on an open $U$ of $X$, then
the form $\omega$ of a connection $\nabla$ in the frame $s$ is defined by the matrix relation 
\begin{equation}\label{connectionform}
\nabla s= \omega s,
\end{equation}
 where $s$ is regarded as a ``column vector''. Then the curvature form with respect to $s$ is the $2$-form
\begin{equation}\label{Curvatureform}
d\omega-\omega\wedge\omega.
\end{equation}

\smallskip

Not every vector bundle $F$ on the manifold $X$ admits an integrable connection.
By the Frobenius theorem, the existence of a flat connection  on ${\mathcal F}$ is equivalent to the fact that   $F$ admits a family of local frames, whose domains cover $X$ and such that the corresponding transition functions are constant; i.e., $F$ is a flat vector bundle 
\cite[page 5]{Koba}.

Conversely, if a vector bundle admits integrable connections, then it is flat. 
In this case,
   each integrable connection 
    defines, via the holonomy, a representation of the fundamental group $\pi_1(X)$ in 
   the group of automorphisms of the fiber \cite[Chapter I, \S 2]{Koba}.
	
	On the other hand, each finite dimensional representation of $\pi_1(X)$ determines a flat complex vector bundle $V$ on $X$ equipped with a integrable connection $\nabla$. Moreover, equivalent representations of $\pi_1(X)$ determine isomorphic 
  $(V,\,\nabla)$ pairs, and conversely.



\smallskip

Given an     {\it integrable} connection $\nabla$ on the holomorphic vector bundle $F$, 
 we have the following complex
 associated to $\nabla$
\begin{equation}\label{Complex^.}
  \Omega^{\bullet}_{X} \otimes_{{\mathcal O}_X}{\mathcal F} :\;\; {\mathcal F}\overset{\nabla}{\longrightarrow}
  \Omega^1_{X}  \otimes_{{\mathcal O}_X} {\mathcal F} \overset{\nabla^1}{\longrightarrow}   \Omega^2_{X}  \otimes_{{\mathcal O}_X} {\mathcal F}     \overset{\nabla^2}{\longrightarrow}\dots
 \end{equation}

One can consider the cohomology sheaves ${\mathcal H}^j\big(  \Omega^{\bullet}_{X} \otimes_{{\mathcal O}_X}{\mathcal F}  \big)$. In particular, 
\begin{equation}\label{R^0pi}
{\mathcal H}^0\left(  \Omega^{\bullet}_{X} \otimes_{{\mathcal O}_X}{\mathcal F}  \right)
 ={\rm Ker}(\nabla)=:{\mathcal K}.
 \end{equation} 
 Thus, we can associate to each integrable connection the cohomology  groups   
 $H^p\left(X,\,{\mathcal H}^q( \Omega^{\bullet}_X\otimes_{{\mathcal O}_X}{\mathcal F}) \right).$

\smallskip

 We will prove that 
  the Poincar\'e's lemma holds for the complex (\ref{Complex^.}).
Given 
$$\alpha\cdot s:=\sum_c\alpha^c\otimes  s_c\in {\rm Ker}(\nabla ^k)(U),$$ 
we need to prove the existence of a form 
$\beta\cdot s\in \big(\Omega^{k-1}_{X}  \otimes_{{\mathcal O}_X} {\mathcal F}\big)(\tilde U) $,
with $\tilde U\subset U$, such that $\nabla^{k-1}(\beta\cdot s)=\alpha\cdot s$ on $\tilde U$.

From the condition $0=\nabla(\alpha\cdot s)$, it follows
\begin{equation}\label{omegawedge}
d\alpha+(-1)^k\alpha\wedge\omega=0.
\end{equation}
 And the ``unknown'' $\beta$ must satisfy the equation.
\begin{equation}\label{alpha=dbeta}
\alpha=d\beta+(-1)^{k-1}\beta\wedge\omega.
\end{equation} 
Differentiating this equation and using (\ref{omegawedge}), one arrives to the integrability condition
$$\beta\wedge d\omega=\beta\wedge\omega\wedge\omega.$$
 This condition is satisfied, since the curvature form (\ref{Curvatureform}) vanishes. 
 Hence,  the Poincar\'e's lemma holds for the complex 
 $  \Omega^{\bullet}_X \otimes_{{\mathcal O}_X} {\mathcal F}$,
and
\begin{equation}\label{HqDeRham}
{\mathcal H}^q\left(\Omega_X^{\bullet} \otimes_{{\mathcal O}_X} {\mathcal F} \right)=0,\quad \text{for}\; q>0.  
 \end{equation}

The kernel ${\mathcal K}$ of $\nabla$ is the local system consisting of the parallel sections of ${\mathcal F}$. It is  a locally free ${\mathbb C}_X$-module. Thus,  any section  of ${\mathcal K}$ over an open set $W$ contained in an open  connected    $U\subset X$ can be uniquely extended to a section of ${\mathcal K}$  over $U$.

  Let ${\mathfrak U}=\{U_j\}$ be a good covering of $X$ such that ${\mathcal K}|_{U_j}$ is a constant sheaf. Then the 
	$\check{\rm C}$ech 
	cohomology 
	$\check H(X,\,{\mathcal K})=\check H({\mathfrak U},\,{\mathcal K})$. By the above uniqueness of the extension, given a cocycle in 
	$\check{\mathcal C}^q({\mathfrak U},\,{\mathcal K})$, $q\geq 1$, is a coboundary. Hence,
	\begin{equation}\label{Cechcoho}
	 H^q(X,\,{\mathcal K})=0,\;\;\; \text{for}\; q>0.
	\end{equation}

	The spectral sequence 
	$$E_2^{p,q}=H^p\left(X,\, {\mathcal H}^q( \Omega_X^{\bullet} \otimes_{{\mathcal O}_X} {\mathcal F} )\right)$$
	abuts to hypercohomology ${\mathbb H}^*(X,\,  \Omega_X^{\bullet} \otimes_{{\mathcal O}_X} {\mathcal F} ).$ By (\ref{HqDeRham}) together with (\ref{R^0pi}),
	$${\mathbb H}^q\left(X,\,    \Omega_X^{\bullet} \otimes_{{\mathcal O}_X} {\mathcal F} \right) \simeq H^q(X,\,{\mathcal K}).$$
	Here we have a proof of the following known result.
	 \begin{Prop}\label{HpC^bullet}
	The hypercohomology of the
	complex (\ref{Complex^.}) is  
	$$	{\mathbb H}^p\left(X,\,  \Omega_X^{\bullet} \otimes_{{\mathcal O}_X} {\mathcal F}  \right)=  \begin{cases} \Gamma(X,\,{\mathcal K}),\quad\text{for}\,\;p=0 \\
	0,\quad\hbox{for}\,\;p>0. 
	\end{cases}$$
	\end{Prop}

In summary, given a flat gauge field $\nabla$, the cohomological content of the     complex (\ref{Complex^.}) reduces to the space of parallel sections of $F$.


\smallskip
                                                  
\subsubsection{Riemann-Hilbert-Deligne correspondence.}\label{RHD}
 Let $Y$  be a submanifold of the complex manifold $X$. By ${\mathcal O}_X[Y]$ we denote the sheaf of meromorphic functions on $X$, 
 which are holomorphic on $X\setminus Y$ and that have poles in $Y$. Let  ${\mathcal G}$ be  a coherent ${\mathcal O}_X[Y]$-module. A linear ${\mathbb C}_X$-linear map
$$\nabla:{\mathcal G}\to \Omega^1_X\otimes_{{\mathcal O}_X}{\mathcal G}$$
 satisfying
$$\nabla(f\sigma)= df\otimes\sigma +f\nabla\sigma,$$
with $f$ and $\sigma$ are sections of ${\mathcal O}_X[Y]$ and ${\mathcal G}$ (resp.) 
is called a   connection on ${\mathcal G}$  meromorphic on  $Y$ \cite{Malgrange}. We will say that   that $\nabla$ is flat when
$$\left[\nabla_v,\,\nabla_{v'}\right]=\nabla_{[v,\,v'] },$$
 for arbitrary vector fields $v,v'$.   

According to  Fuchs' theory, the solutions to a linear ordinary differential equation with regular singularities have a moderate growth. The regularity condition admits a translation to the meromorphic connections (see \cite[Chapter 5]{Hotta}).

When studying some topological phases, we will find examples of meromorphic  connections that are of the type described below. We denote by $\Omega^k_X(\log Y)$ the sheaf of $k$-forms on $X$ with logarithmic poles along $Y$. If $j:X\setminus Y\to X$ is the inclusion,  $\Omega^k_X(\log Y)$ is the subsheaf of $j_*\Omega^k_X$ such the stalk at $y\in Y$
$$\Omega^k_X(\log Y)_y=\left\{\alpha\in (j_*\Omega^k_X)_y\,|\, \varphi\alpha\in \Omega^k_{Xy},\; \varphi d\alpha\in \Omega^{k+1}_{Xy} \right\},$$
 where $\varphi$ is an equation of $Y$ near of $y$. The 
  connections referred  are connections on vector bundles over $X$, such that the connection forms $\omega\in\Omega^k_X(\log Y)$.

 The Deligne's version of the Riemann-Hilbert correspondence gives an identification of the flat connections on vector bundles which have  logarithmic poles along a divisor $D$ of $X$,   and the representations of the fundamental group $\pi_1(X\setminus D)$. Essentially, that correspondence is defined via the holonomy around the loops in $X\setminus D$ \cite{Malgrange} \cite[Chapter II]{Deligne}. 
 

\subsubsection{$D$-modules}\label{Sub:D-modules}

 By ${\mathcal D}_X$ we denote the sheaf of differential operators on $X$. The canonical sheaf ${\mathcal K}_X$ is a right ${\mathcal D}_X$-module \cite[page 9]{Kashiwara}. On  $D^b({\mathcal D}_X)$, the bounded derived category of left ${\mathcal D}_X$-modules, is   defined the de Rham functor ${ DR}$ \cite[page 103]{Hotta}
$$
DR:D^b({\mathcal D_X})\to D^b ({\mathbb C}_X),\qquad {\mathcal M}^{\bullet}  \mapsto {\mathcal K}_X\otimes_{{\mathcal D}_X}^L {\mathcal M}^{\bullet},$$ 
 where $ D^b({\mathbb C}_X)$ is the bounded derived category   of ${\mathbb C}_X$-modules.

Given a ${\mathcal D}_X$-module ${\mathcal M}$, one can consider 
the following complex
 $(\Omega_X^{\bullet}\otimes_{{\mathcal O}_X}{\mathcal M},\,\delta^{\bullet})$ with
 \begin{equation}\label{deldeRham}
\delta^k(\alpha\otimes m)=d\alpha\otimes m+\sum_{i=1}^n(dx_i\wedge\alpha)\otimes (\partial_{x_i}\cdot m),
\end{equation}
 where $\alpha$ is a homolorphic $k$-form 
  and $(x_1,\dots,x_n)$ are local coordinates on $X$. This complex shifted $n$ places to the left represents ${DR}{\mathcal M}$ 
	\cite[page 104]{Hotta}.

 Given a locally free sheaf ${\mathcal F}$ over $X$, an integrable connection $\nabla$ on it determines  a left ${\mathcal D}_X$-module structure on ${\mathcal F}$, in which the action of a vector
 field $v$ on a section $\sigma$ is given by 
\begin{equation}\label{vcdot_sigma}
v\cdot\sigma=\nabla_{v}\sigma.
\end{equation}
Hence, in this case, 
the operator  $\delta^k$ coincides with $\nabla^k$ 
defined in (\ref{nabla_k}) 
and ${ DR}{\mathcal F}$ is represented by the complex  (\ref{Complex^.}) shifted $n$ positions to the left.

\smallskip	
	
 Associated to  a coherent  ${\mathcal D}_X$-module ${\mathcal M}$ is its characteristic variety ${\rm Ch}({\mathcal M})$ 
\cite[page 17]{Kashiwara}. It is an analytic subvariety of the cotangent bundle $T^*X$, and its dimension satisfies ${\rm dim}( {\rm Ch}({\mathcal M}))\geq n$. The ${\mathcal D}_X$-module is called {\it holonomic} if ${\rm dim}( {\rm Ch}({\mathcal M}))= n$. 

  If ${\mathcal M}$ is a holonomic ${\mathcal D}_X$-module, there exists a Whitney stratification of $X$, $X=\sqcup_a X_a$ such that 
  \begin{equation}\label{CharStrat}
  {\rm Ch}({\mathcal M})=\bigsqcup _{a}T^*_{X_a}X,
  \end{equation} 
  where  $T^*_{X_a}X$ is the conormal bundle of $X_a$.
  
       It is easy to show that if ${\mathcal F}$ is a locally free ${\mathcal O}_X$-module endowed with an integrable connection, then the characteristic variety of the corresponding  ${\mathcal D}_X$-module is $T_X^*X$, i.e. the zero section of $T^*X$. The converse is also true. Thus, one has the equivalence between the category of locally free ${\mathcal O}_X$-modules endowed with integrable connections, and the category of coherent ${\mathcal D}_X$-modules such that its characteristic variety is $T^*_XX$ \cite[page 61]{Hotta}.
       
			\smallskip
			
	The sheaf ${\mathcal D}_X$ of differential operators on $X$ has a natural filtration ${\mathcal F}_k({\mathcal D}_X)$, defined by the order of the operators. A filtration of a coherent ${\mathcal D}_X$-module ${\mathcal M}$ is a sequence ${\mathcal F}_k({\mathcal M})$ of submodules of ${\mathcal M}$,
  such that 
  $$ \bigcup_k {\mathcal F}_k({\mathcal M})={\mathcal M},  \quad {\mathcal F}_k({\mathcal M})\subset{\mathcal F}_{k+1}({\mathcal M}),\quad 
 {\mathcal F}_j({\mathcal D}_X) {\mathcal F}_k({\mathcal M})\subset{\mathcal F}_{k+j}({\mathcal M}). $$
 We denote by ${\rm Gr}({\mathcal M})$ the respective graded module
 $$ {\rm Gr}({\mathcal M})=\bigoplus_k {\mathcal F}_k({\mathcal M})/{\mathcal F}_{k-1}({\mathcal M}).$$

   Let ${\mathcal I}$ be the defining ideal of the characteristic variety  ${\rm Ch}({\mathcal M})$. The holonomic  ${\mathcal D}_X$-module ${\mathcal M}$ is called {\it regular}, if it locally has a coherent filtration  satisfying ${\mathcal I}{\rm Gr}({\mathcal M})=0$
	\cite[page 101]{Kashiwara}.		
			
    \smallskip 
 
 A sheaf ${\mathcal S}$ of vector spaces on $X$ such that the stalk ${\mathcal S}_x$ is finite dimensional for any $x$ is said 
to be {\it constructible}, if  there exists a stratification $\{Y_a\,|\, a=1,\dots,t \}$  of $X$, such that the restrictions ${\mathcal S}|_{Y_a}$ are locally constant sheaves.
 By $D^b_c({\mathbb C_X})$  we denote the full subcategory of $D^b({\mathbb C}_X)$  consisting of the objects whose cohomology modules are constructible.

 The Kashiwara constructibility theorem   asserts that for any 
 holonomic ${\mathcal D}_X$-module
${\mathcal M}$, the de Rham complex is cohomologically constructible. More precisely, 
 the sheaves 
 \begin{equation}\label{Constr}
{\mathcal H}^j(DR{\mathcal M})|_{X_a},
 \end{equation}
 where the $X_a$'s are the members of the stratification in (\ref{CharStrat}),
 are locally constant \cite[page 89]{M-S(2)} \cite[page 114]{Hotta}.

In fact, according to the Riemann-Hilbert correspondence \cite[page 103]{Kashiwara},  
the functor $DR$
 defines an imbedding  functor from the category of regular  holonomic ${\mathcal D}_X$-modules in the category $D^b_c({\mathbb C}_X)$.
More precisely, the functor is an equivalence from the regular holomorphic ${\mathcal D}_X$-modules to the perverse sheaves on $X$.  Hence,  ${\mathcal H}^j(DR{\mathcal M})|_{X_a}$ determines a representation $\rho_{j,a}$ of the respective fundamental homotopy group. 
     
By $D^b_{rh}({\mathcal D}_X)$ one denotes the full subcategory of the derived category of $D^b({\mathcal D}_X)$, consisting of those objects whose cohomology modules are regular holonomic ${\mathcal D}_X$-modules. 
 The Riemann-Hilbert correspondence states the equivalence between $D^b_{rh}({\mathcal D}_X)$ and $D_c^b({\mathbb C}_X)$ \cite[page 174]{Hotta}. Thus, each object of $D^b_{rh}({\mathcal D}_X)$ defines a collection of local systems on analytic subspaces of $X$.



\section{Flat gauge fields}\label{S:Flat}


 Let $F$ be a Hermitian {\it flat} vector bundle over $X$ 
of rank $m$. By $\langle \cdot\,,\,\cdot\rangle$ we denote the Hermitian metric on $F$. Let $\nabla$ be a connection on $F$ compatible with  $\langle\,,\,\rangle$; that is,
$$d\langle\sigma,\,\sigma'  \rangle=\langle\nabla\sigma,\,\sigma' \rangle+\langle\sigma,\,\nabla\sigma'  \rangle,$$
for any sections $\sigma,\sigma'$.

If  $s=(s_1,\dots,s_m)$ be a local unitary frame of $F$ defined on an open $U$ of $X$ and $\omega$ is the connection form in $s$,
$$0=d\langle s_i,\,s_j\rangle=\Big\langle\sum_a \omega_{ia}s_a ,\,s_j\Big\rangle + \Big\langle s_i ,\,\sum_b\omega_{jb}s_b\Big\rangle=\omega_{ij}+\bar\omega_{ji}.$$ Thus, $\omega$ is a $1$-form 
${\mathfrak u}(m)$-valued.


\begin{Lem} If  $\nabla$ is a flat connection  compatible with $\langle\,,\,\rangle$, then there is a family $\{s,s',\dots\}$ of local unitary frames, whose domains cover $X$ and such that  $\nabla s=0,\,\nabla s'=0,\dots .$
\end{Lem}

{\it Proof.} Given a local unitary frame $\tilde s$, we are looking for a matrix $B$ of functions, such that $\nabla s=0$,
where $s:=B\cdot\tilde s$.  Denoting by   $\tilde\omega$ the form of $\nabla$ with respect $\tilde s$ 
$$0=\nabla s=(dB+B\tilde\omega)\tilde s.$$
That is,   $B$ must satisfy
\begin{equation}\label{dB+btildeomega}
dB+B\tilde\omega=0.
\end{equation} 
The reasoning used for  equation (\ref{alpha=dbeta})  can be applied to (\ref{dB+btildeomega}).  Thus, there exists a local function $B$ satisfying (\ref{dB+btildeomega}). As  $\tilde\omega$ is ${\mathfrak u}(m)$-valued, since $\tilde s$ is unitary, $B$ is a ${\rm U}(m)$-valued function. Thus, $s$ is a unitary local frame satisfying $\nabla s=0$.
\qed
\smallskip

For the local unitary frames $s, s'$ such that $\nabla s=0=\nabla s'$,    the transition function $g$, defined by $g  s'=s$, is a {\it constant}  ${\rm U}(m)$-valued function. 

A section $\zeta$   of $F$ over $U$ can be written as $\zeta=a\cdot s$, where $a$ is a map $a:U\to {\mathbb C}^m$. The section $\zeta$ is  parallel if
\begin{equation}\label{parallel}
a\omega +da=0.
\end{equation}
The parallel transport defined by (\ref{parallel}) can be expressed in terms of the above constant  transition functions. If $\gamma:[0,\,1]\to X$ is a path in $X$, let $t_0=0<t_1,\dots, <t_k=1$ be a partition of interval, such that $\gamma([t_i,\,t_{i+1}])\subset U_i$,
where $U_i$ is the domain of a local unitary frame $s^{(i)}$ satisfying $\nabla s^{(i)}=0$. If $g_i$ is the constant transition function $s^{(i+1)}=g_is^{(i)}$, the parallel transport along $\gamma$ is the element $g_k g_{k-1}...g_0\in {\rm U}(m)$. This transport determines
the corresponding holonomy   representation $\rho$ of the fundamental group of $X$ 
$$\rho:\pi_1(X) \to {\rm U}(m).$$

Another flat connection  $\nabla'$ on ${\mathcal F}$ compatible with $\langle\,,\,\rangle$ is    equivalent to $\nabla$, if there exists an automorphism $T$ 
of ${\mathcal F}$ such that
\begin{equation}\label{gaugeEquiv}
\nabla\circ T=T\circ\nabla'.
\end{equation}
 In this case,  the automorphism $T$ is defined by a smooth map 
$h:X\to {\rm U}(m)$, and   by (\ref{gaugeEquiv}), the connection form of $\nabla'$ in the frame $s$ is
$$\omega'=dh\,h^{-1}+h\omega h^{-1}.$$

The equation for the parallel transport with respect to $\nabla'$ is $a' \omega' +da'=0$. Hence, if $a'$ is solution for this equation, then $ a'h$ is solution of (\ref{parallel}).
Given    $\epsilon\in \pi_1(X,\,x_0)$ we denote by
 $A,\,A'\in{\rm U}(m)$ the respective holonomies around $\epsilon$. By the preceding result $A'=  h(x_0)Ah^{-1}(x_0)$. That is, the corresponding holonomy representations $\rho,\rho'$ are equivalent.


\smallskip

Conversely, let $\lambda:\pi_1(X)\to{\rm U}(m)$ be a repsentation of the homotopy group of the connected complex manifold $X$.   $p:\tilde X\to X$, will  denote the universal covering of $X$. Let $H:=(\tilde X\times {\mathbb C}^m)/\sim$, where  
$$(\tilde x,v)\sim \big(\epsilon\tilde x,\,\lambda(\epsilon)v\big),$$
for $\epsilon\in \pi_1(X)$. We set $\pi:H\to X$ for the obvious projection.

Let $\kappa:U\to \tilde X$ a local section of $p$. For $i=1,\dots, m$ we define 
$$s_i=\left[\kappa(x),\,e_i\right]\in \pi^{-1}(x),$$
with $e_i=(0,\dots,1,\dots,0)\in{\mathbb C}^m$. Then $s=(s_1,\dots, s_m)$ defines a local trivialization of $H$. If $s,s'$ are trivializations
of this type, defined by the local sections $\kappa$ and $\kappa'$ of $p$,   there exists $\epsilon \in \pi_1(X)$ with $\epsilon\kappa=\kappa'$. Hence, the transition function between $s$ and $s'$ is a {\it constant} function $\lambda(\epsilon)$. Then we can define a flat connection $\Hat\nabla$ on $H$ by declaring $\Hat\nabla s=0,$ $\Hat\nabla s'=0,\dots$. This construction of $(H,\Hat\nabla)$ from the representation $\lambda$ is the inverse of the preceding one, in which we have constructed the representation $\rho$ from the pair $(F,\nabla)$.  
Thus, one has the following proposition.

\begin{Prop}\label{P:FlatBundle}
Let $F$ be a Hermitian flat vector bundle over $X$. There is a bijective correspondence between the equivalence classes of integrable connections on $F$ compatible with the Hermitian structure, and the equivalence classes of unitary representations of $\pi_1(X)$.
\end{Prop}

When $X$ is an oriented Riemannian compact manifold and $F$ is a Hermitian vector bundle, one defines the action functional (Yang-Mills functional) on the connections compatible with Hermitian structure:
\begin{equation}\label{Y-M_functional}
\nabla\mapsto {\rm YM}(\nabla)=\int_X || K_\nabla\wedge\star K_{\nabla}||\,
\end{equation}
where $\star$ is the Hodge operator and $||\, .\,||$ is the norm defined by the Hermitian metric  $\langle\,,\,\rangle$ of $F$ \cite[page 417]{Hamilton} \cite[page 44]{Moore}. The connections on which the above functional takes stationary values are the Yang-Mills fields. If $F$ admits flat gauge fields,  these connections minimize the functional ${\rm YM}$. These fields are the vacuum states of the corresponding Yang-Mills theory \cite[page 447]{Hamilton}.
From    Proposition \ref{P:FlatBundle}, one deduces the following corollary.

\begin{Cor}\label{Cor-Vacuum} Let $F$ be a Hermitian flat vector bundle of rank $m$ over the oriented connected  Riemannian manifold $X$. Then  the vacuum states of the theory are in bijective correspondence with the classes of equivalent representations   of $\pi_1(X)$ in  ${\rm U}(m)$. 
\end{Cor}

	
	
	\begin{Prop}\label{pi1=Z} Under  the hypotheses of Corollary \ref{Cor-Vacuum}.
	If $\pi_1(X)$ is an infinite cyclic group, 
	then the space of vacuum states of the corresponding Yang-Mills theory can be identified with     the quotient space 
$({\rm U}(1))^m/\sim$, where
$$\left(z_1,\dots,z_m\right)\sim \left(z_{\tau(1)},\dots,z_{\tau(m)}\right),$$
$\tau$ being a permutation of $\{1,\dots, m\}$.
\end{Prop}

{\it Proof.} A representation $\rho$ of  $\pi_1(X)$ in ${\rm U}(m)$  is determined by the value of $\rho$ at a 
	fixed   generator  $o$ of $\pi_1(X)$. Hence, two representations $\rho$ and $\rho'$ are equivalent iff there exists $S\in{\rm U}(m)$ such that   $\rho'(o)=S\rho(o)S^{-1}$. 
	
	On the other hand, any matrix of ${\rm U}(m)$ is conjugate with a matrix of the maximal torus $({\rm U}(1))^m$. Moreover, two matrices of this maximal torus are conjugated iff the have the same spectrum. 
	Thus, the proposition follows from Corollary \ref{Cor-Vacuum}.
	
	\qed

\begin{Prop}\label{pi1=Z2}  Under  the hypotheses of Corollary \ref{Cor-Vacuum},
	if $\pi_1(X)={\mathbb Z}/2{\mathbb Z}$,   
 then the space of vacuum states is a set of $m+1$ elements.
\end{Prop}
{\it Proof.}
 The irreducible representations of the symmetric group $S_k$ of $k$ elements correspond to the conjugacy classes, which in turn are in bijective correspondence with    the    partitions of the number $k$.  Thus, ${\mathbb Z}/2{\mathbb Z} $ has only two irreducible representations $r,\, r'$, which are of dimension $1$, since the group is abelian. The 
 vacuum  states, i.e. the representations of $\pi_1(X)$ in ${\rm U}(m)$, are the direct sums 
of the form $(\oplus_i r)\oplus(\oplus_j r')$, with $i,j\in{\mathbb Z}_{\geq 0}$ and $i+j=m$. Thus,  the proposition follows.

\qed


\section{Topological phases}\label{S:TopologFases}
The topological phases, which appear in the wave function of a particle when it moves in flat gauge fields,  can be interpreted in the context of the Deligne's theory. Next, we consider some of these cases.

\medskip

 {\sc Aharonov-Bohm efect.} Let us consider an infinitely long solenoid in ${\mathbb R}^3$, with radius $R$ and whose axis is the $x^3$ coordinate axis. Denoting by $r$ the usual cylindric coordinate, the magnetic field created by the solenoid is 
 $$\vec B=\begin{cases} B\Hat e_3,\quad \hbox{for}\; r\leq R \\
  0,\quad \hbox{for}\;r>R
  \end{cases}
  $$
$B$ being a constant. A continuous potential vector is
$$\vec A=\begin{cases} \tfrac{B}{2}(-x^2,\,x^1,\, 0),\quad \hbox{for}\;r\leq R \\
   \tfrac{BR^2}{2r^2}(-x^2,\,x^1,\, 0), \quad \hbox{for}\;r>R
  \end{cases}
  $$
The shift in the phase of an electron around the circle $C$, $r=a>R$ positively oriented, is up to a constant factor
$$\oint_C\vec A\cdot d\vec l=\Phi,$$
where $\Phi=B\pi R^2$ is the magnetic flux through the solenoid. In fact, the phase factor in Stoney units is ${\rm e}^{-i\Phi}$  \cite[page 28]{Bohm_et_al}.

The corresponding strength Faraday form for the field created by the solenoid in $r<R$ is
$$K=\frac{1}{2i}\sum_{i,j,k=1}^3B_idx^j\wedge dx^k=-iBdx^1\wedge dx^2,$$
and $K=0$ on $r>R$.
Introducing the complex coordinate $z=x^1+ix^2$, the $1$-form 
$$\tilde \alpha=\frac{-B}{2}\bar zdz$$ satisfies $d\tilde \alpha=K$ on $|z|\leq R$.
 On the other hand, the $1$-form 
$$\alpha=\frac{-\Phi}{2\pi z}dz,$$
is closed in $|z|>R$. Moreover, 
 $\tilde \alpha=\alpha$ in $|z|=R$.

 As the relevant quantity in the Aharonov-Bohm effect is $\Phi:=B\pi R^2$, we can imagine the magnetic field confined along the $x^3$-axis, and such that its flux trough a disc with center on the $x^3$-axis and orthogonal to this axis is $\Phi$. Motived by this idealization, we introduce the following notation
$X:={\mathbb C}$, $D:=\{0\}$ and 
${\mathcal F}:={\mathcal O}_X[D]$. 
 In this setting, $\alpha$
 can be regarded as a meromorphic flat connection, with logarithmic singularity at $z=0$,  on the ${\mathcal O}_X[D]$-module ${\mathcal F}$. 

 As $\oint_C \alpha=-i\Phi$, according to Deligne's result, that connection determines a local system on ${\mathbb C}\setminus 0$, which is the one that corresponds to the representation of ${\mathbb Z}$ in ${\rm U}(1)$ determined by $-i\Phi$.

\smallskip

Let $S_1,\dots,S_r$ be a collection of solenoids in the space, and denote by $\pi$ an affine  plane in the space   that contains none of the solenoids. By fixing an origin in $\pi$, a complexification of the resulting vector space defines a complex coordinate $z$ on $\pi$. Let us denote by
$\Phi_1,\dots,\Phi_r$ the fluxes of the corresponding solenoids.
If $z_1,\dots,z_r$ are the complex coordinates of  the intersections of these ideals solenoids with that plane $\pi$, then the meromorphic form 
$$\frac{-1}{2\pi}\sum_j\frac{\Phi_j}{z-z_j}\,dz$$  
determines a local system on ${\mathbb C}\setminus\{z_1,\dots,z_r\}$.
 The effect of the set    of solenoids on an electron moving on the plane $\pi$ around a closed curve $\gamma$, which encloses only the points $z_{j_1},\dots, z_{j_k},$ will be a phase factor 
$${\rm exp}\Big(-i\sum_{a=1}^k\Phi_{j_a}\Big).$$  

\medskip


{\sc Aharonov-Casher effect.} The above discussion can be generalized to   a particle which carries a spin-like variable. An example is the case of a non relativistic neutron moving around a charged wire. The corresponding phase change is the well-known Aharonov-Casher 
effect \cite[page 254]{C-J}.

A uniformly charged wire along the $x^3$ coordinate axis in ${\mathbb R}^3$ produces the  electric field
\begin{equation}\label{vecE}
\vec E=\frac{q}{2\pi r^2}\left( x^1\Hat{e}_1+x^2\Hat e_2  \right),
\end{equation}
$q$ being the static density of charge and $r^2=(x^1)^2+(x^2)^2$.

The spin angular moment of the neutron can be written $S=\sum_{a=1}^3S^a\tau_a$, with 
$$\tau_a=\frac{-i}{2}\,\sigma_a,$$
and where the $\sigma_a$'s are the Pauli's matrices. The structure constants of $\mathfrak{su}(2)$ relative to
 the basis $\{\tau_a\}_{a=1,2,3}$ are the Levi-Civita symbols $\epsilon_{ikj}$.  

Denoting $\vec S:=(S^1,\,S^2,\,S^3),$ the equation of motion for $S$ is (\cite[page 365]{Jackson})
\begin{equation}
\label{JackSk}\frac{d}{dt}\vec S=-\lambda\vec S\times(\vec v\times \vec E),
 \end{equation}
where $\lambda$ is a constant, and $\vec v$ is the velocity of  neutron. Assuming that the neutron is moving on the plane $x^1,x^2$, from (\ref{vecE}), it follows
 $$\frac{d}{dt}S^1=-\lambda\left(E_2v^1-E_1v^2\right)S^2,\;\;\frac{d}{dt}S^2=\lambda\left(E_2v^1-E_1v^2\right)S^1,\;\; \frac{d}{dt}S^3=0.$$

On the other hand, the Wong equation for the motion of $S$ is 
\begin{equation}\label{WonSk}
\frac{d}{dt}\,S^k=\sum A^i_{\mu}v^{\mu}\epsilon_{ikj}S^j,
\end{equation}
where the $\mathfrak{su}(2)$-valuated $1$-form $A=\sum A^i_{\mu}dx^{\mu}\tau_i$ is the potential gauge. 

Using (\ref{JackSk}) and (\ref{WonSk}), we deduce that one can take as potential 
 $$A=\left(A^3_1dx^1+A^3_2dx^2\right)\tau_3,$$
 with $A^3_1=-\lambda E_2$ and $A^3_2=\lambda E_1$.  That is,
 \begin{equation}\label{Lambdapoten}
A=\frac{\Lambda}{r^2}\left(x^2dx^1-x^1dx^2\right)\tau_3,
\end{equation}
$\Lambda$ being $-\frac{\lambda q}{2\pi}$. 

The phase factor accumulated by the spin of neutron when it completes an anticlockwise  circle $C$ in the plane $x^1,x^2$ around the origin is
\begin{equation}\label{CaserBig}
\rm{exp}\left(\oint_CA\right)= \rm{exp}(i\pi \Lambda\sigma_3)=\begin{pmatrix} {\rm e}^{i\pi\Lambda} & 0 \\
  0 & {\rm e}^{-i\pi\Lambda} \end{pmatrix}.
	\end{equation}
	
	\smallskip
	Next, we show an interpretation of this result in the context of Deligne's theory.
	On ${\mathbb R}^2\setminus\{(0,\,0)\}$ 
	$$\frac{x^1dx^1+x^2dx^2}{r^2}=d\log r.$$
	Hence, one can add to the potential form (\ref{Lambdapoten}) the exact form 
	$$\frac{i\Lambda\left(x^1dx^1+x^2dx^2\right)}{r^2}\,\tau_3.$$
	Setting $z=x^1+ix^2$, the new potential form can be written as
	$$\alpha:=\frac{i\Lambda\, dz}{z}\,\tau_3.$$
	This form is a meromorphic flat connection on the sheaf ${\mathbb C} \tau_3\otimes_{\mathbb C} {\mathcal O}_X[D]$, where $X={\mathbb C}$ and $D=\{0\}$.
	The regular meromorphic connection $\alpha$ corresponds, in the Deligne's theory,  to the representation of $\pi_1({\mathbb C}^{\times})$ which associates to
	the loop
	$\gamma:t\in[0,\,2\pi] \to {\rm e}^{it}$ the value  
	${\rm exp}(\int_{\gamma}\alpha)$, which is  equal to the phase factor (\ref{CaserBig}).

  \medskip
  
  {\sc Holonomy and the Wong equation.}
	Let $X$ be a complex manifold and $D$ a normal crossing divisor of $X$.
  That is, if $p$ is a point of $D$, there exist   $x_1,\dots,x_n$  coordinates around $p$ such that $D$ is defined by $x_1\cdot\cdot\cdot x_r=0$. 
  
  Let $G$ be a matrix Lie group, subgroup of ${\rm U}(m).$  By  $B$ we denote  a meromorphic  $1$-form on $X$ with values in ${\mathfrak g}$ such that
  \begin{enumerate}
  \item $B$ is holomorphic on $X\setminus D$ and has   poles along $D$,
  \item    $dB-B\wedge B=0,$ on $X\setminus D$ (see (\ref{Curvatureform})).
  \end{enumerate}
   We also assume that $B$ 	has logarithmic singularities along  $D$. That is, if the local expression of $B$   around $p$ is $\sum_{\mu}B_{\mu}dx^{\mu}$, with $B_{\mu}=(B_{\mu, ij})_{i,j=1,\dots, m}$
	 the functions
   $ x_{\mu} B_{\mu, ij}$ (for $1\leq\mu\leq r$) and $B_{\mu, ij}$ (for $r<\mu\leq n$) are holomorphic. 
  
  Let us consider a particle carrying a like spin variable $I$ which takes values in the Lie algebra ${\mathfrak g}$. We assume that the particle is moving on the manifold $X$,  where  the gauge field $B$   is present. 
  If its trajectory is the   curve $\gamma$,
\begin{equation}\label{curvex}
t\in[0,\,1]\mapsto x(t)\in X,
 \end{equation}
	the variation of $I$ along $\gamma$ satisfies the Wong equation \cite[page 53]{Bala}
	\begin{equation}\label{WongEqu}
	\frac{d}{dt}I=\left[B(\Dot x(t)),\, I(t)\right].
	\end{equation}
  In terms of a basis  $\{\tau_a\}$ of ${\mathfrak g}$,   $I=\sum_a I^a \tau_a$ and $B(\Dot x)=\sum_b B^b(\Dot x)\tau_b$; thus, the above equation   gives rise to known expression
	 \begin{equation}\label{WonIa}
	 \frac{d}{dt}I^a=\sum_{bc}f^a_{bc}B^b(\Dot x)I^c,
	 \end{equation}
	where the $f^a_{bc}$ are the structure constants of ${\mathfrak g} $ relative to the basis $\{\tau_a\}$. 
	
	Since (\ref{WongEqu}) is a Lax equation, by the isospectral property, the eigenvalues of $I(t)$ are independent of $t$. Moreover, if $\vec v_0$ is an eigenvector of $I(0)$  associated to  the eigenvalue $\lambda$, then an eigenvector of $I(t)$ associated to $\lambda $ can be obtained  by propagating $\vec v_0$, that  is, solving the equation
		\begin{equation}\label{propagation}
		\frac{d}{dt}\vec v=B(\Dot x(t))\vec v(t),\qquad\vec v(t=0)=\vec v_0.
		\end{equation}
  Denoting by ${\mathcal R}_B $  the  resolvent of this equation, one has   $\vec v(t)={\mathcal R}_B(t)\vec v_0.$ 

Thus, if $\{w_i\}$ is a basis of eigenvectors of $I(0)$, with $I(0)w_i=\lambda_iw_i$, then 
$$I(1)S w_i=\lambda_i S w_i,$$
where $S={\mathcal R}_B(1).$ That is,
 \begin{equation}\label{I(1)I(0)}
 I(1)=SI(0)S^{-1}.
 \end{equation}
 That is, assumed that $\gamma$ is a closed curve,  the initial value $I(0)$ of the variable $I$
 changes according to (\ref{I(1)I(0)}), after the movement along the  curve $\gamma$.

\smallskip

  We denote by  $E$ the trivial vector  bundle $X\times{\mathbb C}^m$ over $X$, then the form $A:=-B$ defines a flat connection $\nabla$   on 
  ${\mathcal O}_X[D]\otimes_{{\mathcal O}_X} {\mathcal E}$ with logarithmic singularities along $D$.

  Denoting by $\{u_i\}_{i=1,\dots,m}$ the global frame of $E$ defined by the canonical basis of ${\mathbb C}^m$, a section $\xi=\sum_i\xi_iu_i$ of $E$ is parallel along the above curve $\gamma$ if
	\begin{equation}\label{paralleltransport}
	\frac{d}{dt}\,\vec\xi=-A(\Dot x(t))\vec\xi(t),
	\end{equation}
	where $\vec\xi$ is the column vector $(\xi_1,\dots,\xi_m)$.
	We will denote  by ${\mathcal R}_{-A}(t)$
	the corresponding resolvent. 
	In the case that the curve $\gamma$ is closed,   then the holonomy around $\gamma$ is ${\rho}(\gamma):={\mathcal R}_{\small{-A}}(1)\in G\subset {\rm U}(m).$
	
	According to the Deligne's version of the Riemann-Hilbert correspondence, 
	\begin{equation}\label{rhoRepresentation}
	\rho:\pi_1(X\setminus D)\to G\subset{\rm Aut}({\mathbb C}^m)
	\end{equation} 
	is the representation of $\pi_1(X\setminus D)$  determined by the flat connection $\nabla$.

	\smallskip

As $B=-A$, $S={\mathcal R}_{-A}(1)$, so   from (\ref{I(1)I(0)}) one deduces the following matrix equation
\begin{equation}\label{ChangeI}
I(1)=\rho(\gamma)I(0)(\rho(\gamma))^{-1}.
 \end{equation} 
Thus, shifts in the variable $I$, after the movement along the closed curves on $X\setminus D$, can be collected  in the  
 following representation 
$$\tilde\rho={\rm Ad}\circ\rho:\pi_1(X\setminus D)\rightarrow  {\rm Aut}({\mathfrak g}),$$
where ${\rm Ad}$ is the adjoint representation of ${\mathfrak g}$. 
 Hence, we get the following proposition.
	\begin{Prop}\label{Prop:Wong}
	 Let $G$ be a Lie subgroup of ${\rm U}(m)$ and   $B$ be a flat   gauge field, that takes values in ${\mathfrak g}$,  with logarithmic singularities along the divisor $D$ of $X$. Let $p$ be a $G$-Yang-Mills particle, that carries a spin like variable  $I$ with values in ${\mathfrak g}$, moving in the   field $B$. Denoting by 
	  $\tilde\rho$   the representation of $\pi_1(X\setminus D)$ determined by the phase factor of $I$, and by $\rho$ the holonomy representation defined by the connection $-B$, then
$$\tilde\rho={\rm Ad}\circ\rho.$$
  \end{Prop}
  

\section{Flat gauge fields on holomorphic sheaves}\label{S:FlatConnections}


\subsection{Gauge field on a locally free $B$-brane} 
If $({\mathcal F},\,\nabla)$ is an integrable connection, the complex (\ref{Complex^.}) represents $DR{\mathcal F}[-n]$. As we said
 ${\rm Ch}({\mathcal F})=T^*_XX$. From Proposition \ref{HpC^bullet}, it follows that the set of local systems (\ref{Constr})  reduces to ${\mathcal K}$, the local system of parallel sections. The corresponding representation of $\pi_1(X)$ is equivalent to the one of Proposition \ref{P:FlatBundle}, when $F$ is an Hermitian bundle and $\nabla$ is compatible with the Hermitian structure. 


\smallskip

Let us assume that $({\mathcal F}^{\bullet},\,d^{\bullet})$ is a bounded complex of finitely generated holomorphic {\it locally free sheaves} on the manifold $X$. It is natural to define a gauge field on the corresponding $B$-brane as a family $\{\nabla^{(\bullet)}\}$ of connections, where $\nabla^{(i)}$ is a connection on ${\mathcal F}^i$  such that the following diagrams are commutative  
  



\begin{equation}\label{cuadrado}
\xymatrix{
   \Omega_X^k \otimes_{{\mathcal O}_X}  {\mathcal F}^i \ar[r]^{\nabla^{(i)k} }\ar[d]_{ {\rm id}  
	\otimes   d^{i}} &  \Omega_X^{k+1 }  \otimes_{{\mathcal O}_X} {\mathcal F}^i    \ar[d]^ { {\rm id}  
	\otimes   d^{i}} \\
  \Omega_X^k  \otimes_{{\mathcal O}_X}  {\mathcal F}^{i+1}     \ar[r]^{{\nabla^{(i+1)k} }  } & \Omega_X^{k+1}  \otimes_{{\mathcal O}_X} {\mathcal F}^{i+1}    \,.}
\end{equation}
The existence of these sequences of compatible connections is proved in \cite{B-B}.

Given   such a family $\{\nabla^{(i)}\}$ of {\it flat} connections, one has the double complex
$\left({\mathcal C}^{\bullet,\bullet},\,\nabla^{(\bullet)},d^{\bullet}\right)$, where
$${\mathcal C}^{p,q}=  \Omega_X^p  \otimes_{{\mathcal O}_X}  {\mathcal F}^q.$$
 We set ${\mathcal T}^{\bullet}$ for the total complex. Equipped each ${\mathcal F}^i$ with   the ${\mathcal D}_X$-structuture defined by the connection $\nabla^{(i)}$ according to (\ref{vcdot_sigma}), one has the object ${\mathcal F}^{\bullet}$ of the category 
$D^b({\mathcal D}_X)$ and $DR{\mathcal F}^{\bullet}$ is represented the complex ${\mathcal T}^{\bullet}[n]$.

We will determine the hypercohomology groups ${\mathbb H}^k(X,\,{\mathcal T}^{\bullet})$ in order to study the cohomological content of the flat gauge field on the brane. 

 The spectral sequence $E_2^{p,q}=H^p(X,\, {\mathcal H}^q\left({\mathcal T}^{\bullet})\right)$ abuts to ${\mathbb H}^{p+q}(X,\, {\mathcal T}^{\bullet})$.
On the other hand, the spectral sequence $\tilde E_2^{p,q}=H^p_{d}H_{\nabla}^q({\mathcal C}^{\bullet,\bullet})$ converges to $H^{p+q}({\mathcal T}^{\bullet})$.

Since the ${\mathcal F}^p$ are {\em locally free sheaves},   the Poincar\'e's lemma holds for the complexes 
$ \Omega_X^{\bullet} \otimes_{{\mathcal O}_X} {\mathcal F}^p    $. 
By (\ref{HqDeRham}) 
$$H^q_{\nabla}\left(  \Omega_X^{\bullet} \otimes_{{\mathcal O}_X} {\mathcal F}^p \right)  =0,$$
 for $q>0$ and any $p$.
Thus, 
 $\tilde E_2^{p,q}=0$ for $q>0$ and
\begin{equation}\label{DefK}
H^0_{\nabla}\left( \Omega_X^{\bullet}   \otimes_{{\mathcal O}_X} {\mathcal F}^p   \right)={\rm Ker}\left({\mathcal F}^p\overset{\nabla ^{(p)}}{\longrightarrow}   \Omega^1_X \otimes_{{\mathcal O}_X} {\mathcal F }^p  \right)=:{\mathcal K}^p.
 \end{equation}  
By the convergence of this spectral sequence, 
\begin{equation}\label{HpT}
H^p({\mathcal T}^{\bullet})=H^p_d({\mathcal K}^{\bullet}).
\end{equation}

On the other hand, each  ${\mathcal K}^i$ is a local system, since its sections are the parallel sections of ${\mathcal F}^i$. Hence, the cohomology sheaves $H^j_d({\mathcal K}^{\bullet})$ are also local systems \cite[page 53]{M-S(2)}, and consequently
$$\check H^i(X,\,H_d^j({\mathcal K}^{\bullet}))=\begin{cases} 0,\quad \hbox{for}\; i>0  \\
                              \         \Gamma\left(X,\,H_d^j({\mathcal K}^{\bullet})\right),\quad \hbox{for}\; i=0\; .
																			\end{cases} $$
Thus, from (\ref{HpT}) it follows
$E_2^{p,q}=0$ for $p>0$ and $E_2^{0,q}=\Gamma(X,\,H^q_d({\mathcal K}^{\bullet})).$ By the convergence of this spectral  sequence, one deduces the following proposition.
\begin{Prop}\label{Hypercoho}
If   the ${\mathcal F}^p$  are holomorphic locally free sheaves and the connections $\nabla^{(p)}$ are flat, then the hypercohomology
$${\mathbb H}^p(X,\,{\mathcal T}^{\bullet})=\Gamma\left(X,\,H^p_d({\mathcal K}^{\bullet})\right),$$
where ${\mathcal K}^i$ is the locally constant sheaf defined in (\ref{DefK}).
\end{Prop}

Assumed that there exist  families of flat connections on the brane $({\mathcal F}^{\bullet},\,d^{\bullet})$, 
since ${\mathcal T}^{\bullet}[n]$ represents $DR{\mathcal F}^{\bullet}$,  the local systems $H^p_d({\mathcal K}^{\bullet})$ correspond to those   considered in (\ref{Constr}) (in the present case the stratification of $X$ is the trivial one).  
Each  $H^p_d({\mathcal K}^{\bullet})$  determines  a representation  of the homotopy group $\pi_1(X)$, and these representations give the  phase factors of the respective Aharonov-Bohm effect.

 \subsection{Flat gauge fields on a $B$-brane.}
 In this section we propose a   definition  of a flat gauge field on a $B$-brane. 
 
 An integrable connection on a locally free sheaf ${\mathcal F}$ defines  a parallel transport, and this transport in turn determines the connection. Furthermore, the equation $\nabla \sigma=0$ for the parallelism has the following properties
\begin{enumerate}
\item  It  is a overdetermined equation, so the space of solutions is finite dimensional.
\item It is a regular differential equation in Fuchs' sense,   even when $\nabla$ has logarithmic singularities along a divisor. Thus,   
its solutions have a moderate growth.
 \end{enumerate}
On the other hand, given a flat gauge field on   ${\mathcal F}$, the action (\ref{vcdot_sigma})  defines a ${\mathcal D}_X$-module structure on  ${\mathcal F}$.
    
    The above argument leads one to think that certain singular ${\mathcal D}_X$-structures on a 
		coherent ${\mathcal O}_X$-module ${\mathcal E}$,	
		can be considered as  flat gauge fields on this $B$-brane. The holonomic regular ${\mathcal D}_X$-modules satisfy  the properties that are the translation of (1) and (2) into the context of the $D_X$-modules.

		
		\smallskip
		
		Let $Y$ be a submanifold  of $X$. 
		The sheaf ${\mathcal O}_X[Y]$ has been defined in Section
		\ref{RHD}. Given a coherent ${\mathcal O}_X$-module ${\mathcal E}$, we denote by
		${\mathcal E}^Y$ the ${\mathcal O}_X[Y]$-module
			$$ {\mathcal E}^Y:={\mathcal O}_X[Y]\otimes_{{\mathcal O}_X}{\mathcal E}.$$
			A flat  connection $\nabla$ on $ {\mathcal E}^Y$, meromorphic along $Y$ (see Section \ref{RHD}) defines a holonomic $ {\mathcal D}_X$ structure on 
		$ {\mathcal E}^Y$ \cite[page 140]{Hotta}. Such a connection is called regular when the ${\mathcal D}_X$-module  $ {\mathcal E}^Y$ is regular. 
		
		For $Y=\emptyset$, we set ${\mathcal E}^Y:={\mathcal E}$ and   set ${\mathcal O}_X[\emptyset]={\mathcal O}_X$. With this convention  the flat connections on ${\mathcal E}$ can also be considered as flat regular connections on ${\mathcal E}^Y$ meromorphic along $Y$.
		We adopt the following definition.

	\begin{Def}
	A  {\it flat gauge field on the coherent module} ${\mathcal E}$ is a  flat regular meromorphic connection on $ {\mathcal E}^Y$, where $Y$ is a fixed divisor   of $X$ or $Y=\emptyset$. 
	\end{Def}
	 
	 \smallskip
	 
		By the Riemann-Hilbert correspondence, a flat gauge field determines a perverse sheaf on $X$. Which in turn defines representations $\rho_{j,a}$ of the homotopy groups of the strata $X_a$ of a Whitney statification of $X$ (see Section \ref{Sub:D-modules}). 
		That is, for each $a$ one has representations $\rho_{ja}$ of $\pi_1(X_a)$.  
In this way, given a closed path $\gamma$ in 
$X_a$,
the values of  $\rho_{ja}([\gamma])$ will be  the   phase factors of the corresponding ``Aharonov-Bohm'' effect.

		The singularity set of the coherent ${\mathcal O}_X$-module ${\mathcal E}$ is a closed  analytic  subset $Z\subset X$ with dimension $\leq n-1$ \cite[page 88]{Scheja}.  We assume that $Z$ is a submanifold of $X$,
		then the flat connections on ${\mathcal E}^Z$ are smooth on the points $x\in X$ where ${\mathcal E}_x$ is a free ${\mathcal O}_{Xx}$-module.

\smallskip

Next, we consider the $B$-brane on $X$ defined by the complex $({\mathcal E}^{\bullet},\,d^{\bullet})$ of coherent 
${\mathcal O}_X$-modules.		Given a divisor $Y$ of $X$ (or $Y=\emptyset$), we   construct the complex
$$\left({\mathcal E}^{\bullet Y},\,\tilde d^{\bullet}\right):=\left({\mathcal O}_X[Y]\otimes_{{\mathcal O}_X}{\mathcal E}^{\bullet},\,1\otimes d^{\bullet}\right).$$

We define a {\it flat gauge field} on that brane as a family $\nabla^{(i)}$, where
\begin{itemize}
\item $\nabla^{(i)}$ is a connection over ${\mathcal E}^{iY}$, that is     flat regular and meromorphic along $Y$, and
\item 
 the $\nabla^{(i)}$'s are compatible with the operators $\tilde{d}^{i}$.    
\end{itemize}

Therefore, the flat gauge field on the brane, converts it in an object of the category $D^b_{rh}({\mathcal D}_X)$. Which in turn defines an object of $D^b_c({\mathbb C}_X)$, by the Riemann-Hilbert correspondence. The representations determined by  the corresponding local systems can be interpreted as generalized Aharonov-Bohm phase factors.  
 

 \smallskip

{\sc Aharonov-Bohm phase and perverse sheaves.} As we have seen in Section \ref{S:TopologFases}, the phase factor on the  wave function of an electron that completes a closed curve around the solenoid is ${\rm e}^{-i\Phi}$. We assume that  there exists Aharonov-Bohm effect; that is,  the flux $\Phi\notin 2\pi{\mathbb Z}$. We will interpret this fact in the context of perverse sheaves.

Let $X$ be ${\mathbb C}$,    $Y=\{0\}$ and ${\mathcal E}:={\mathcal O}_X$. With the above notations ${\mathcal E}^Y ={\mathcal O}_X[Y]$. One defines the following left action of the sheaf $\Theta_X$ of holomorphic vector fields on $X$ on ${\mathcal E}^Y$
$$\partial_z\cdot s=\frac{ds}{dz}-\frac{\lambda}{z}\,s,$$
for $s\in {\mathcal O}_X[Y]$, $z$ being the coordinate of ${\mathbb C} $ and $\lambda:=-\frac{\Phi}{2\pi }$.
This action defines a ${\mathcal D}_X$-module structure on ${\mathcal O}_X[Y]$. We denote by ${\mathcal M}$ the corresponding ${\mathcal D}_X$-module. 
 The following complex is the de Rham complex of ${\mathcal M}$
	$$({\mathcal C}^{\bullet}, \,d^{\bullet}):\; \to 0\to{\mathcal C}^{-1}={\mathcal M}\overset{\delta}{\longrightarrow} 
	{\mathcal C}^{0}=\Omega^1_X\otimes_{{\mathcal O}_X}{\mathcal M}\to 0 \to$$
where $\delta(s)=dz\otimes(\partial_z\cdot s)$.

The equation $z\frac{du}{dz}=\lambda u$ has as solution $u=({\rm const})z^{\lambda}$; as
$\lambda\notin{\mathbb Z}$, it defines holomorphic branches on $X\setminus Y$. 
 Thus, the cohomology sheaf ${\mathcal H}^{-1}( {\mathcal C}^{\bullet})$ is the extension by zero at $0$ of the local system on $X\setminus Y$ associated with the multivalued  function $z^{\lambda}$. On the other hand, ${\mathcal H}^{0}({\mathcal C}^{\bullet})$ is trivial. Hence, the cohomology sheaves of the de Rham complex   are constructible with respect to the following stratification of $X$
$$X_0=Y,\; \, X_1=X\setminus Y.$$

Among the local systems ${\mathcal H}^j({\mathcal C}^{\bullet})\big|_{X_a}$, the only non trivial corresponds to $j=-1$ and $a=1$.
 If $g$ is a local section of  ${\mathcal H}^{-1}({\mathcal C}^{\bullet})\big|_{X_1}$, 
then $g(z)$ is equal (up to constant factor) to $z^{\lambda}={\rm exp}(\lambda\log z)$, where $\log z$ is any branch of the logarithm function. The analytic continuation of $g(z)$ along the curve $\gamma=\{{\rm exp}(2\pi it)\,|\,t\in[0,\,1]\}$, transforms 
$g(z)$
into $g(z){\rm exp}(2\pi i\lambda)$. As the homotopy class of $\gamma$ generates $\pi_1(X_{1})$, the   representation $\rho_{-1,1}$ defined by the local system ${\mathcal H}^{-1}({\mathcal C}^{\bullet})\big|_{X_1}$ is determined by
$$\rho_{-1,1}([\gamma])={\rm exp}(2\pi i\lambda)={\rm e}^{-i\Phi},$$
 which is the phase factor in Aharonov-Bohm effect.

Next, we prove that the object ${\mathcal C}^{\bullet}$ of the category $D^b_c({\mathbb C}_X)$ is a perverse sheaf. For $S\in\{X_0,\,X_1\}$ one must to check \cite[Proposition 2.2.2]{BBD}
$${\mathcal H}^j\left(i_S^{-1}{\mathcal C}^{\bullet}\right)=0,\;\; j>-\,{\rm dim}\, S \quad \hbox{and}\quad
 {\mathcal H}^j\left(i_S^{!}{\mathcal C}^{\bullet}\right)=0,\;\; j<-\,{\rm dim}\, S, 
  $$
where $i_S^{-1}$ and $i_S^{!}$ are the inverse image functor and the inverse image with compact functor defined by the inclusion
 $i_S:S\hookrightarrow X$
\cite[Section III.8]{Ge-Ma}.

 Hence, we need only to consider 
the sheaves ${\mathcal H}^0\left(i^{-1}_S{\mathcal C}^{\bullet}\right)$  (when $S=X\setminus Y$) and
 ${\mathcal H}^{-1}\left(i^{!}_S{\mathcal C}^{\bullet}\right)$ (when $S=Y$).  For the first case, as $i^{-1}_S$ is an exact functor
$${\mathcal H}^0\left(i^{-1}_S{\mathcal C}^{\bullet}\right)=i^{-1}_S({\mathcal Coker} (\delta))={\mathcal Coker} (\delta)\big|_{X\setminus Y}=0.$$ 
For the second case, since for any sheaf ${\mathcal A}$, $i^{!}_S{\mathcal A}$ is the sheaf of sections of ${\mathcal A}$ with support at $0$,
then ${\mathcal H}^{-1}\left(i^{!}_S{\mathcal C}^{\bullet}\right)=0$.

\smallskip

{\sc Brane wrapping a subvariety.}
Let  $Z$ be a spin submanifold of $X$. Then the anticanonical line bundle $K_Z^{-1}$ has  square roots \cite[page 396]{L-M}. We denote by
 $K_Z^{-{1}/{2}}$ such a root. If $F\to Z$ is  vector bundle over $Z$, we set
$${\mathcal G}:=i_*\left({\mathcal F}\otimes_{{\mathcal O}_Z} K_Z^{-{1}/{2}}\right),$$
 where $i$ is the inclusion $i:Z\hookrightarrow X$. Thus ${\mathcal G}$ is a  coherent ${\mathcal O}_X$-module, which from the physical point of view is   a $B$-brane on $X$ wrapping $Z$ \cite[page 68]{Aspin} (see also \cite{Katz-Sharpe}).

 If $Y$ is a divisor  of $X$,  
a flat connection on ${\mathcal G}^Y$,   regular and meromorphic along $Y$  defines a representation of 
$\pi_1(X\setminus Y)$, 
 which can be interpreted as an Aharonov-Bohm phase factor in  brane ${\mathcal G}$ endowed with this gauge field. 

 In the case that $X$ is a Calabi-Yau manifold, then the canonical line bundle ${K}_X$ is trivial. Furthermore, if $Z$ is defined by a section of a line bundle $L$ over $X$, by the adjunction formula \cite[page 147]{G-H}
$$K_Z=(K_X\otimes L)|_Z=L|_Z.$$ When $F$ is the trivial line bundle,  ${\mathcal F}={\mathcal O}_Z$, then 
$${\mathcal G}=i_*i^{-1}{\mathcal L}^{-{1}/{2}}={\mathcal L}_Z^{-{1}/{2}}.$$
A flat gauge field 
 on the line bundle 
${\mathcal L}^{-1/2}$  determines by restriction a  gauge field 
on the brane ${\mathcal G}$, and  
 the phase factors, along any loop contained in $Z$, in both fields are the same.




\end{document}